\documentclass[hyper]{JHEP3}
\pdfoutput=1
\usepackage{amsmath}
\usepackage{amssymb,amsfonts,amsxtra,mathrsfs,makeidx,graphics,graphicx,amsthm,bbm}
\usepackage{epsfig}
\usepackage[all]{xy}
\usepackage{feynmf}
\usepackage{slashed}

\newcommand{\CC}{\mathcal{C}}

\newcommand{\CS}{\mathcal{S}}

\newcommand{\CL}{\mathcal{L}}
\newcommand{\CM}{\mathcal{M}}
\newcommand{\CN}{\mathcal{N}}
\newcommand{\CO}{\mathcal{O}}

\newcommand{\CD}{\mathcal{D}}
\newcommand{\CH}{\mathcal{H}}
\newcommand{\CV}{\mathcal{V}}

\renewcommand{\Im}{{\rm Im}}

\newcommand{\SU}{\mathrm{SU}}

\newcommand{\half}{\frac{1}{2}}

\newcommand{\nn}{\nonumber}

\def\p{\partial}

\def\bea{\begin{eqnarray}}
\def\eea{\end{eqnarray}}
\def\be{\begin{equation}}
\def\ee{\end{equation}}
\def\ba{\begin{align}}
\def\ea{\end{align}}
\def\bse{\begin{subequations}}
\def\ese{\end{subequations}}
\newcommand{\ben}{\begin{eqnarray}}
\newcommand{\een}{\end{eqnarray}}

\newcommand{\bem}{\begin{pmatrix}}
\newcommand{\eem}{\end{pmatrix}}

\def\={\;  = \;}
\def\+{\, + \,}

\def\wt{\widetilde}
\def\wh{\widehat}
\def\bar{\overline}

\def\rt2{\sqrt{2}}

\renewcommand{\Im}{\mbox{Im}}

\def\g{\gamma}

\def\a{\alpha}

\def\O{{\Omega}}

\def\ve{\varepsilon}

\def\nv{n_{\rm v}}

\title{Quantum black hole entropy and the holomorphic prepotential of $\CN=2$ supergravity}

\preprint{NIKHEF2013-019}

\author{Sameer Murthy$^\dagger$ and Valentin Reys$^{\dagger\ddagger}$ \\
$^\dagger$ Nikhef theory group, Science Park 105, \\
1098 XG Amsterdam, The Netherlands \\
$^\ddagger$ LPTHE, Universit\'e Pierre et Marie Curie, place Jussieu, \\
F-75252 Paris Cedex 05, France \\

{\tt \email{smurthy, vreys at nikhef dot nl}} }

\abstract{ 
Supersymmetric terms in the effective action of $\CN=2$ supergravity in four dimensions are generically classified into 
chiral-superspace integrals and full-superspace integrals. For a theory of~$\CN=2$ vector multiplets coupled to 
supergravity, a special class of couplings is given by chiral-superspace integrals that are governed by a holomorphic prepotential function. 
The quantum entropy of BPS black holes in such theories depends on 
the prepotential according to a known integral formula. We show, using techniques of localization, that a large 
class of full-superspace integrals in the 
effective action of $\CN=2$ supergravity do not contribute to the quantum entropy of BPS black holes 
at any level in the derivative expansion. 
Our work extends similar results for semi-classical supersymmetric black hole entropy, and 
goes towards providing an explanation of why the prepotential terms capture the exact microscopic 
quantum black hole entropy.
}


\begin{document}

\section{Introduction and summary}

It was proposed in the 1970s by Bekenstein and Hawking that black holes have a thermodynamic entropy 
equal to a fourth of the area of the event horizon in Planck units. This \emph{area-law} is a 
semi-classical formula and holds when the black hole horizon area is very large compared to the Planck 
scale. The quantum entropy of black holes is a generalization of the area-law that 
takes into account the quantum fluctuations of matter and gravitational fields in a black hole. 
The effects of these fluctuations are encoded in corrections to the area-law that 
are suppressed when the area of the horizon in Planck units is infinite.

The fluctuations of massive fields in a black hole background 
can be summarized in a local effective action that includes higher dimension operators in addition to 
the theory of general relativity minimally coupled to matter fields that is universally valid at low energies. The contributions 
of these local higher-dimensional operators to the entropy are taken into account by the extension 
due to Wald~\cite{Wald:1993nt, Iyer:1994ys} of the Bekenstein-Hawking formula.
The quantum fluctuations of light fields, on the other hand, give rise to non-analytic and non-local terms 
in the 1PI effective Lagrangian, and one needs a full functional integral treatment 
to take these effects into account. For supersymmetric black holes, such a treatment was proposed by Sen 
in~\cite{Sen:2008yk, Sen:2008vm}. 
The formal idea is to integrate over all the fields of the gravitational theory with boundary conditions 
set by the $AdS_{2}$ attractor configuration arising in the near-horizon region of the black hole. 

For a class of black holes in string theory in four and five dimensions with 16 or more supersymmetries, 
we can calculate the 
exact microscopic degeneracy of BPS states~$d(Q_{i})$ as a function of the charges~\cite{Dijkgraaf:1996it, Maldacena:1999bp, Banerjee:2007sr, Banerjee:2008ri, Banerjee:2008pu, Dabholkar:2008zy}.
In the limit of infinite charges, the function~$d(Q_{i})$ obeys 
a Cardy-like formula, and the statistical entropy~$S_{\rm micro} \equiv \log(d(Q_{i}))$ agrees with the thermodynamic 
entropy given by the Bekenstein-Hawking area-law~\cite{Strominger:1996sh}.
One can go further and extract the subleading corrections to the leading Cardy-like formula for 
the microscopic entropy~\cite{LopesCardoso:1998wt, LopesCardoso:2000qm, David:2006yn, Gaiotto:2005gf, 
Shih:2005qf, Shih:2005he, Castro:2008ys, Dabholkar:2010rm} (see also~\cite{Sen:2007qy} for a review).
We expect that the degeneracy of states (or more precisely the supersymmetric
index~\cite{Dabholkar:2010rm}) does not change on moving in moduli space\footnote{This is strictly true in the absence of wall-crossing.
The situation is more complicated when there is wall-crossing~\cite{Denef:2007vg}, but there 
has also been progress in finding 
explicit generating functions for the black hole degeneracy in a class of examples with~$\CN=4$ 
supersymmetry~\cite{Dabholkar:2012nd}.}. 
The subleading corrections to the microscopic degeneracy thus act as a check for the quantum corrections to the thermodynamic 
gravitational entropy of the black hole. 
Unlike the leading area-law which is a universal 
formula valid for any black hole in general relativity, the subleading corrections depend crucially on the 
structure of the gravitational theory beyond the leading two-derivative action.

A comparison between the microscopic and thermodynamic entropy including subleading power-law corrections was first performed in~\cite{LopesCardoso:1998wt, LopesCardoso:2000qm} for four-dimensional black holes in $\CN=2$ supergravity coupled to vector multiplets, 
 using a local effective action 
that included four-derivative terms suppressed by two powers of the string scale~$\ell_{\rm s}$ compared to the 
leading universal two-derivative action of supergravity. 
The most general supersymmetric action in such 
a theory of~$\CN=2$ supergravity can be naturally divided into chiral-superspace integrals that are captured 
by the holomorphic prepotential function~$F$~\cite{deWit:1979ug}, and full-superspace integrals, both of which
admit an infinite expansion in~$\ell_{\rm s}$. 
The authors of~\cite{LopesCardoso:1998wt, LopesCardoso:2000qm} considered a four-derivative theory 
that only contained terms of the first type and found that the corrections agreed with the microscopic counting functions.

More recently, a method to sum up all the perturbative quantum contributions to the quantum entropy of supersymmetric black holes, 
including the quantum effects of massless fields, was put forward in~\cite{Dabholkar:2010uh}. The method relies 
on an adaptation of the technique of supersymmetric 
localization~\cite{Witten:1988ze, Witten:1991zz, Schwarz:1995dg, Pestun:2007rz}
 which reduces the full supergravity functional integral to a finite dimensional manifold called the \emph{localization manifold} $\CM_{Q}$. The final formula for the quantum entropy has the following form:
\begin{equation}\label{integralintro}
 \wh W (q, p) = \int_{\mathcal{M}_{Q}}   \exp\big(\CS_{\rm ren}(\phi, q, p) \big) \, [d \mu(\phi)] \, .
\end{equation}
The integrand in this formula is the exponential of the supergravity action evaluated on the localization manifold,  
with a suitable renormalization to get rid of infra-red divergences~\cite{Sen:2008yk}. The measure $[d \mu(\phi)]$ and some 
other details of this formula are presented in~\S2.3.

The authors of~\cite{Dabholkar:2010uh} made a further assumption (as in~\cite{LopesCardoso:1998wt, LopesCardoso:2000qm}) that the supergravity action is fully governed by the holomorphic prepotential~$F$. In this case 
the renormalized action takes the form:
\begin{eqnarray}
 \mathcal{S}_{ren}(\phi, q, p) =  - \pi  q_I   \phi^I + 4 \pi \, \Im \left[ F\Big(\frac{\phi^I+ip^I}{2} \Big) \right] \, ,
\end{eqnarray}
which gives a formula of the type originally conjectured by~\cite{Ooguri:2004zv}.
The prepotential~$F$ can be computed for~$\CN=2$ supergravity theories that arise as Calabi-Yau compactifications of type II string theory using methods of topological string theory~\cite{Antoniadis:1993ze, Bershadsky:1993cx}. 
When the~$CY_{3}=T^{6}$ all the higher genus topological string amplitudes vanish, and the classical cubic 
prepotential is exact at all orders in~$\a'$. In this case, the values for the exponential of the quantum entropy 
agreed with the integer degeneracy predicted for these black holes by string theory to exponential 
accuracy~\cite{Dabholkar:2011ec}. 

These results suggest that the exact quantum entropy for a generic $\CN=2$ supergravity theory coupled to vector multiplets 
is fully captured by the holomorphic prepotential. In other words, although the 
effective action that enters~\eqref{integralintro} may contain an infinite number of 
higher-derivative full-superspace integrals, none of them seem to contribute to the exact quantum entropy. 
This generalizes the corresponding suggestion for the semi-classical\footnote{Here and in the rest 
of the paper, we follow the recent literature in the use of the phrases  ``quantum'' and ``semi-classical'' entropy to 
distinguish if the quantity takes 
into account the effects of massless fields running in loops or not. In this terminology, the semi-classical entropy can 
include the effects of higher-derivative corrections encoded in a local effective action. 
A more clear nomenclature (that is usually used in field-theory contexts) may be to use the phrases ``1PI'' and ``Wilsonian'' entropy.} entropy based on~\cite{LopesCardoso:1998wt, LopesCardoso:2000qm}, 
for which evidence was provided in~\cite{deWit:2010za}. 

In this paper, we shall provide similar evidence for the above statement concerning the non-renormalization of quantum entropy. 
In particular, a large class of full-superspace integrals that can be added to the~$\CN=2$ supergravity action 
can be written down explicitly~\cite{deWit:2010za}. We show that none of these known full-superspace integrals contribute to the full quantum entropy. 

A very brief summary of our method of proof is as follows: the localization manifold~$\CM_{Q}$ is the set of solutions 
of the off-shell BPS equations and is independent of the choice of action, and so the contribution of any new 
term to the quantum entropy is 
controlled by its value on the points of the localizing manifold. 
We show here that the full-superspace integrals vanish when evaluated on the localizing manifold. 
In addition, the measure and the electric charges do not change under the 
addition of such terms. Taken together, these facts imply that the functional integral for quantum black hole 
entropy in four-dimensional $\CN=2$ supergravity theories coupled to vector multiplets is independent of 
such full-superspace integrals in the effective Lagrangian.

The plan of the rest of the paper is as follows. In~\S2, we briefly summarize the classical black hole attractor solution
and the quantum entropy function formalism, and we review the method of localization as applied to the calculation of the quantum entropy function in supergravity. In~\S3, we present the class of full-superspace integrals that we 
consider in this paper, and we review the result of~\cite{deWit:2010za}, namely that they do not contribute to the semi-classical
entropy. This result is a necessary background for our quantum result that we present in~\S4.
In~\S5, we present a short discussion of our results and of further extensions. 
We display some details of our calculations and of the Euclidean continuation that we use in two appendices.

\section{Quantum black hole entropy and localization}

In this section, we first briefly review the BPS black hole solutions in the~$\CN=2$ supergravity theory that we are interested in. Next we review the concept of quantum entropy as applied to these black holes. 
We then summarize the computation of the exact quantum entropy of these black holes using the localization formalism.

\subsection{Semi-classical black hole entropy}

We are interested in a theory of~$\CN=2$ supergravity coupled to vector fields. We work in the formalism 
of conformal $\CN=2$ supergravity coupled to~$\nv+1$ vector multiplets~\cite{deWit:1980tn}.
This theory has a local superconformal algebra that extends the local Poincar\'e superalgebra,
and is gauge-equivalent to~$\CN=2$ Poincar\'e supergravity. The local dilatation invariance can be 
gauge-fixed using one of the vector multiplets called the compensating multiplet. Upon gauge-fixing the 
extra symmetries of the superconformal theory, 
we get the~$\CN=2$ Poincar\'e supergravity with the canonical Einstein-Hilbert term for the vielbein. 

The main advantage of this formalism is that the supersymmetries are realized off-shell, and they do 
not need to be modified even when the action of the theory is modified, e.g.~by adding higher-derivative 
terms to the Lagrangian.
This will be crucial to us when we use localization to compute the functional integral for black hole entropy.
We shall present only the aspects that are relevant to us in this paper, and refer the reader 
to the original references and the review~\cite{Mohaupt:2000mj} for more details on the formalism.

The \emph{Weyl multiplet} in the conformal supergravity contains the following independent fields:
\be\label{Weylfields}
{\bf W} \= \left( e_{\mu}^{a}, \,  \psi_{\mu}^{i}, \, b_{\mu}, \, A_{\mu}, \CV_{\mu \, j}^{\, i},  \, T_{\mu\nu}^{ij}, \, \chi^{i}, \, D \right) \, .
\ee
There are also other fields in the multiplet that are composite fields built out of the above fields. 
In the two-derivative gauge-fixed Poincar\'e theory, the field $e_{\mu}^{a}$ is the vielbein, and 
the $T_{\mu\nu}$ tensor is an auxiliary field without kinetic term. These two fields will play an important 
role in our discussion.

The independent fields of the \emph{vector multiplet} are
\be\label{Vectorfields}
{\bf X}^{I} \= \left( X^{I}, \, \O_{i}^{I}, \, A_{\mu}^{I}, \, Y^{I}_{ij}  \right) \, , 
\ee
where $X^{I}$ is a complex scalar, the gaugini $\O^{I}_{i}$ are an $\SU(2)$ 
doublet of chiral fermions, $A^{I}_{\mu}$ is a vector field, and $Y^{I}_{ij}$ are an $\SU(2)$ triplet of 
auxiliary scalars.

In this theory, we are interested in black hole solutions that preserve one half of the supersymmetries.  
They carry electric and magnetic charges $(q_{I},p^{I})$, $I=0,1,\cdots \nv$, and 
interpolate between fully supersymmetric asymptotically flat space and the near-horizon 
$AdS_{2} \times S^{2}$ region. The near-horizon region is a fully supersymmetric solution of the 
theory in its own right, and in the low energy limit, it can be decoupled from the environment and 
studied on its own. We parameterize the  $AdS_{2} \times S^{2}$ as follows:
\bea
\label{metric}
ds^2 &=& v\left[-(r^2-1)dt^2 + \frac{dr^2}{r^2-1}\right] + v\left[d\psi^2 + \textnormal{sin}^2(\psi)d\phi^2\right], \cr
&&\widehat{F}^I_{rt} = e^I_*, \quad \widehat{F}^I_{\psi\phi} = p^I\textnormal{sin}\psi, \quad T^-_{rt} = vw \, .
\eea
Here we have shown the metric and the field strengths of the $\nv + 1$ gauge fields sitting in the vector multiplets
that are relevant in the solution. The tensor field~$T^{-}_{\mu\nu}$ is a component of the auxiliary tensor~$T_{\mu\nu}^{ij}$
that is part of the off-shell graviton multiplet.  This field plays a central role throughout our 
analysis, and we shall discuss it in more detail below. The tensor~$T$ appears in 
most of the equations through the linear combinations~$T_{\mu\nu}^{-}=T_{\mu\nu}^{ij} \, \ve_{ij}$
and its complex conjugate $T_{\mu\nu}^{+}=T_{\mu\nu\, ij} \, \ve^{ij}$, where~$\ve_{ij}$ is the invariant 
tensor of~$SU(2)$.

The complex scalar fields $X^{I}$ of the vector multiplets are determined completely in terms of the 
fluxes by the full-BPS conditions~\cite{LopesCardoso:1998wt, LopesCardoso:2000qm}:
\be\label{attractor}
\widehat{F}_{ab}^{+ I} \= \frac{1}{4} X^I_* \, T^+_{ab} \, , \qquad \widehat{F}_{ab}^{- I} \= \frac{1}{4}\bar{X}^I_* \, T^-_{ab} \, .
\ee
For our solution~\eqref{metric}, we have:
\be
e_*^I + ip^I - \frac{1}{2}X_*^Iv\bar{w} = 0\, , \qquad v=\frac{16}{w\bar{w}}\, .
\ee
The electric fields~$e_{*}^{I}$ are determined in term of the charges~$q_{I}$ as a Legendre transform:
\be \label{Legendre}
\frac{\p \CL^{\rm eff}(e_{*}^{I})}{\p e_{*}^{I}} \= q_{I} \, .
\ee
where $\CL^{\rm eff}$ is the local effective Lagrangian evaluated on the full-BPS configuration~\eqref{metric}, \eqref{attractor}. 
This is the well-known attractor solution in the context of fully supersymmetric black holes.
The function~$\CL^{\rm eff}$ depends on the parameters~$v,e_{*}^{I},\ldots$ in the solution~\eqref{metric}, and 
the Wald-entropy of this black hole is found by extremizing the function with respect to 
its arguments~\cite{Sen:2008yk}.

\subsection{Quantum black hole entropy}

The quantum black hole entropy is defined as a  functional integral over all the fields of the supergravity theory. 
As in standard quantum field theory, this functional integral is defined in Euclidean signature. Since we are 
dealing with curved spacetimes, the Euclidean continuation is more subtle than the usual one in flat space. 
We present some details of this Euclidean continuation in Appendix~\ref{Euclidean}. The Wick rotation on the 
bosonic fields of the classical black hole solution can be effectively carried out by 
the change of variable $t \rightarrow iu$ in the metric~\eqref{metric}.
The fields~$T_{\mu\nu}^{\pm}$ which were complex conjugates in Minkowski signature should be thought of 
as independent fields in Euclidean signature. A similar comment holds for all complex quantities like the 
self-dual components of the field strengths as well as the complex scalars~$X^{I}$.

Quantum mechanically, the $AdS_{2}$  functional integral is defined by summing over all field
configurations which asymptote to these attractor values with the fall-off 
conditions~\cite{Sen:2008yk, Sen:2008vm, Castro:2008ms}:
\bea\label{asympcond}
d s^2 &=& v \left[ 
\left(r^2+\CO(1)\right) d\theta^2+ \frac{dr^2}{r^2+\CO(1)}  \right]\  . \nonumber \\
X^{I} &= &X^{I}_{*} + \CO(1/r)\ ,\qquad
A^I = -i  \, e_{*}^{I} (r -\CO(1) ) d\theta\ .
\eea
The other massive fields asymptote to zero, as is consistent with their classical equations of motion 
near the boundary. 

The functional integral for the partition function is weighted by the exponential of the 
Wilsonian effective action at some fundamental scale defining the theory, such as the string scale.
To make the classical variational problem well-defined, it is necessary to add a boundary term $-i q_{I} \int A^{I}$ 
to the action. 
With this boundary term, the quantum partition function can be naturally interpreted as the expectation 
value of a Wilson line inserted at the boundary
\be\label{qef}
W (q, p) = \left\langle \exp[-i \, q_I \oint_{\theta}  A^I]  \right\rangle_{\rm{AdS}_2}^{finite}\ . 
\ee 
Note that the~$AdS_{2}$ boundary conditions fix all the (electric and magnetic) charges in the theory,
and naturally lead to a microcanonical ensemble.  
The superscript in the above expression refers to the fact that the action of the theory is divergent
due to the infinite volume of $AdS_{2}$, and one therefore needs to regularize it. 
This is done by putting a cutoff $r_{0}$ on the $AdS_{2}$ geometry so that the proper length of the boundary 
scales as $2 \pi \sqrt{v} r_{0}$.   Since the classical action is an integral of a \textit{local} Lagrangian, 
it  scales as ${S_1 r_0 + S_0 + \CO(r_0^{-1})}$. The linearly divergent part can now be subtracted 
by a boundary counter-term, and this procedure sets the origin of energy in the boundary theory. After this renormalization 
we can take  the cutoff to infinity to obtain a finite functional integral  weighted by the exponential of the 
finite piece $S_0$.  This finite piece is a functional of all fields and contains arbitrary higher-derivative terms,
and it is referred to as the renormalized action $S_{ren}$.

A one-loop evaluation of the functional integral~\eqref{qef} for supersymmetric black holes was done in~\cite{Sen:2011ba, Banerjee:2011jp}, 
and the leading logarithmic corrections were successfully matched to the microscopic predictions.
Even a preliminary reading of these papers allows us to appreciate the technical power
used in computing these one-loop corrections. This direct method of computing logarithmic corrections
is applicable in a wide variety of black holes, including non-supersymmetric ones. On the other hand, 
for supersymmetric solutions, the method of supersymmetric localization allows us to sum up the contributions 
from all orders of perturbation theory at one shot. We now turn to a brief review of this method.

\subsection{Computation of quantum entropy using localization}

We review the computation of the quantum entropy~\eqref{qef} of our black hole solutions
using localization~\cite{Dabholkar:2010uh}. One begins by picking a supersymmetry~$Q$
that is realized off-shell in the theory, and that squares to a compact $U(1)$ symmetry. 
One then adds a deformation to the effective action in~\eqref{qef} that is~$Q-$exact, so that the 
functional integral is independent of the deformation.
One then evaluates the functional integral at a convenient point in the deformation space, typically
such that the evaluation reduces to a semi-classical evaluation over a drastically reduced field-space. 
We refer to~\cite{Pestun:2007rz} for a detailed exposition of this method in the context of supersymmetric field theory.

For the conformal supergravity theory 
that we consider, the supersymmetry variations of the gravitini and gaugini fields are: 
\bea\label{gravitinovar}
\delta\psi^i_\mu  & = & 2 \CD_{\mu} \epsilon^i
+\CV_\mu^i{}_j \, \epsilon^j-\frac{1}{4}\gamma^{\rho\nu} \, T_{\rho\nu}^{ij} \, \gamma_\mu \epsilon_j 
-\gamma_\mu \eta^i\, , \\
\label{gauginovar}
\delta \Omega_i^I & = & 2\gamma^\mu D_{\mu} X^I  \epsilon_i+Y^I_{ij} \, \epsilon^j
+\sigma^{\mu\nu}\mathcal{F}^{I-}_{\mu\nu}\varepsilon_{ij} \, \epsilon^j +2X^I\eta_i\, , 
\eea
with 
\be
\mathcal{F}_{\mu\nu}^{I} \equiv \widehat{F}_{\mu\nu}^{I} - 
\Big( \ve_{ij} \overline{\psi}^{i}_{[\mu} \g_{\nu]} \O^{j I} 
+ \ve_{ij} \overline{X}^{I}  \overline{\psi}^{i}_{[\mu} \psi^{j}_{\nu]} 
+ \frac{1}{4}\overline{X}^I T_{\mu\nu}^{ij} \, \varepsilon_{ij} 
+ {\rm h.c.} \Big) \, . 
\ee 
Here~$\epsilon_{i}$ and~$\eta^{i}$ are the parameters of the regular supersymmetry and the conformal 
supersymmetry transformations, respectively. We use the notation that $D_\mu$ is the
covariant derivative covariantized with respect to all the conformal symmetries, while $\mathcal{D}_\mu$
is covariantized with respect to all the conformal symmetries \emph{except} the special conformal boosts
with gauge field $f_\mu^a$~\cite{Mohaupt:2000mj}.

In the geometry~\eqref{asympcond}, we have the generator $L_{0}$ which is the $U(1)$
rotation on the $AdS_{2}$, and another generator $J_{0}$ which is one of the rotations on the $S^{2}$. 
We pick a supercharge that obeys $Q^{2} = L_{0} - J_{0}$ \cite{Dabholkar:2010uh}. 
With this set up, the first step in the localization program is to find all solutions to the equation 
\be\label{qpsi}
Q \, \psi_{\a} =0 \, , 
\ee
where $\psi_{\a}$ runs over all the fermions of the theory. The space of solutions to this equation is called 
the localization manifold~$\CM_{Q}$. 
In the context of~$\CN=2$ conformal supergravity, the complete localization manifold was found 
in~\cite{Gupta:2012cy} and is described as follows.

When the field strength of the $SU(2)$ R-symmetry gauge field $\CV_{\mu \, j}^{\, i}$ (that lives in the graviton 
multiplet \eqref{Weylfields}) is set to zero, the full set of bosonic solutions to the 
localization equations in $\CN=2$ off-shell supergravity coupled to $\nv$ vector multiplets is labelled 
by $\nv+1$ real parameters. 
These parameters label the size of fluctuations of a certain shape 
(fixed by supersymmetry) of the conformal mode of the metric and of the scalars in the $\nv$ vector 
multiplets, and can be taken to be the values of these $\nv+1$ fields at the center of $AdS_{2}$. 
Using the dilatation gauge symmetry of the theory, one can trade the conformal mode of the metric 
for the scalar of the compensating vector multiplet. We set the metric of $AdS_{2} \times S^{2}$
to have unit determinant, and the scalar fields of the vector multiplet have the solution:
\be \label{scalars}
X^I = X^I_* + \frac{w}{4} \frac{C^{I}}{r}, \quad\quad \bar{X}^I = \bar{X}^I_* + \frac{\bar{w}}{4} \frac{C^{I}}{r} \, , 
\qquad I = 0\ldots n_v\, .
\ee
These fluctuations are half-BPS solutions, and they are off-shell. 
They are supported by the auxiliary fields in the vector multiplets:
\be \label{auxfields}
Y_1^{I,1} = -Y_2^{I,2} = \frac{w\bar{w}}{8} \frac{C^{I}}{r^{2}} \, .
\ee
The rest of the fields in the solution remain unchanged with respect to the fully BPS $AdS_{2} \times S^{2}$
solution \eqref{metric}. 
Note that we have included explicit factors of $\frac{w}{4}$ and $\frac{\bar{w}}{4}$ that scale under 
the local dilatation. One can choose a gauge~$w=\bar w =4$ that brings the determinant of the 
metric~\eqref{metric} to unity, but keeping this scale factor manifest is useful in what follows.

An important point to note at the end of the first step is that the localization manifold~$\CM_{Q}$ is universal in that 
it is independent of the choice of the action, since the supersymmetry 
variations~\eqref{gravitinovar}, \eqref{gauginovar} are defined completely in the off-shell theory.

The next step is to evaluate the effective action of the supergravity theory on the localizing 
solutions and correctly integrate over the localizing manifold. The integral has the classical 
induced measure from the supergravity field space, as well as the one-loop determinant of the 
deformation action coming 
from integration over the (non-supersymmetric) directions orthogonal to the localizing manifold in field space:
\begin{equation}\label{integral}
 \wh W (q, p) = \int_{\mathcal{M}_{Q}}   \exp\big(\CS_{\rm ren}(\phi, q, p) \big)
 \, Z_{\rm det}\,  [d\phi] \, ,
\end{equation}
where we have indicated the classical induced measure as~$[d\phi]$ and the one-loop determinant as $Z_{det}$. 
We have displayed this formula in the introduction, wherein we wrote the product of these two factors 
as the full measure~$d \mu(\phi)$. The hat above refers to the fact that only smooth supergravity configurations 
are allowed in this functional integral, while there could be other configurations that are only smooth in the full string 
theory, such as orbifolds, that do contribute to the quantum entropy~\cite{Banerjee:2008ky, Murthy:2009dq}.

In~\cite{Dabholkar:2010uh}, this integral was computed  in the $\CN=2$ supergravity \emph{assuming} that the effective renormalized action $S_{\rm ren}$ only contains chiral-superspace integral terms that are governed by a holomorphic function~$F$ of the vector fields and the Weyl-squared multiplet. With this assumption, and defining the new variables 
\be\label{ephi}
\phi^I := e_{*}^I+2 C^I \ ,
\ee
the renormalized action takes the form:
\begin{eqnarray} \label{Sren}
 \mathcal{S}_{ren}(\phi, q, p) =  - \pi  q_I   \phi^I + \mathcal{F}(\phi, p)\, ,
\end{eqnarray}
with
\begin{equation} \label{freeenergy2}
\mathcal{F}(\phi, p) = - 2\pi i \left[ F\Big(\frac{\phi^I+ip^I}{2} \Big) -
 \bar{F} \Big(\frac{\phi^I- ip^I}{2} \Big) \right] \, .
 \end{equation} 
 
As mentioned in the introduction, this formula was then applied in~\cite{Dabholkar:2011ec} to 
an $\CN=2$ truncation of $\CN=8$ string theory, wherein the microscopic degeneracy of BPS states is known exactly. 
In this case, the prepotential~\eqref{freeenergy2} entering the integral formula is the classical cubic prepotential.  
With some further technical assumptions\footnote{The main 
assumptions are that the hypermultiplets and gravitini multiplets decouple from the vector multiplets in our  
computation, and that the one-loop determinant of the localization action can depend only on the off-shell 
fluctuation of the graviton, and not those of the~$\nv$ physical vector multiplets. Both these issues 
will not affect our conclusions in this paper.}, the quantum entropy for $\CN=8$
black holes could be completely solved, and the answer coming from~\eqref{integral} agreed with the integer microscopic degeneracy to exponential accuracy (see Table 2 in~\cite{Dabholkar:2011ec}). 

The success of this formula points to a non-renormalization theorem of the quantum entropy computed using 
the prepotential. Namely, it seems like full-superspace integrals in the effective action do not contribute to the quantum 
entropy of supersymmetric black holes. In the rest of the paper, we shall provide evidence 
in support of this non-renormalization theorem. In the next section, we shall review the evidence for the 
non-renormalization of the semi-classical entropy, and in~\S4, we shall present new results for the 
non-renormalization of the quantum entropy.

\section{Full-superspace integrals and the semi-classical entropy}

In this section, we review the construction of a large class of full-superspace integrals that can be built in a theory of 
$\CN=2$ supergravity coupled to $\CN=2$ vector multiplets. This is done using the technology of the so-called \emph{kinetic multiplet}~\cite{deWit:1980tn}. We then review the fact that the semi-classical 
black hole entropy does not change on adding these full-superspace terms to the effective action. 
These results were first reported in~\cite{deWit:2010za} which we follow. 
We will suppress fermionic terms in what follows since we are interested in purely bosonic configurations.

\subsection{A large class of full-superspace integral Lagrangians}

Constructing the $\CN=2$ supersymmetric Lagrangians of various matter fields coupled to supergravity is quite an 
intricate technical task. 
The coupling of a chiral multiplet $\Phi$ to supergravity through a chiral-superspace integral was worked out in 
the early days~\cite{deWit:1980tn}: 
\be \label{chiralcoup}
S \= \int d^{4}x \, \CL  \= \int d^{4}x \, d^{4} \theta \, \varepsilon \, \Phi \, , 
\ee
where~$\varepsilon$ is an appropriately defined chiral superspace measure. 
This basic result was then adapted and modified to construct the coupling of vector 
multiplets (by writing the vector multiplet as a reduced chiral multiplet), and to construct higher-derivative terms 
(by considering a holomorphic function $F$ of chiral multiplets as a chiral multiplet itself).
Since~$\theta$ has a Weyl weight~$1/2$, the coupling~\eqref{chiralcoup} is consistent only if the 
superfield~$\Phi$ has weight 2 (so that the action has weight zero).

The same technique can be further modified to construct full-superspace integrals. 
The idea is to construct a kinetic multiplet out of an anti-chiral multiplet, which involves four covariant 
$\bar{\theta}$-derivatives, i.e.~$\mathbb{T}(\bar{\Phi}) \propto \bar{D}^4\bar{\Phi}$. This means 
that~$\mathbb{T}(\bar{\Phi})$ contains up to four space-time derivatives, so that the expression 
\be 
\label{kinetic_int}
\int d^4\theta \, d^4\bar{\theta}\;\Phi \, \bar{\Phi} \approx \int d^4\theta\;\Phi \, \mathbb{T}(\bar{\Phi})
\ee
corresponds to a usual higher-derivative coupling Lagrangian. Here we are being slightly schematic and we have 
not shown the superspace measure.

The field~$\Phi$ and $\bar \Phi$ entering the expression~\eqref{kinetic_int} can be composite fields 
built out of the basic field content of the theory, and can very well be two independent fields. We use 
this fact later in \S4. A more subtle point concerns the nature of the composite 
field~$\bar \Phi$ entering this expression~\cite{BerDan}. In what follows, we shall assume that~$\bar \Phi$
is a physical field that is a local functional of the fluctuating fields of the theory.

From the above expression, one sees that the operator~$\mathbb{T}$ increases the Weyl weight by 2, and 
so the superfield~$\Phi$ should have Weyl weight~$w=0$ in order for the coupling to be consistent. 
For a chiral multiplet~$\Phi$ with components~$(A,\Psi_i,B_{ij},F^-_{ab},\Lambda_i,C)$, 
the Lagrangian~\eqref{kinetic_int} is (see Eqn.~(4.2) of~\cite{deWit:2010za}):
\begin{align} \label{KinLag}
e^{-1}\mathcal{L} =&\,
  4\,\mathcal{D}^2 A\,\mathcal{D}^2\bar A
  + 8\,\mathcal{D}^\mu A\, \big[R_\mu{}^a(\omega,e) -\tfrac13
  R(\omega,e)\,e_\mu{}^a \big]\mathcal{D}_a\bar A + C\,\bar C
  \nonumber \\[.1ex]
  &\,
   - \mathcal{D}^\mu B_{ij} \,\mathcal{D}_\mu B^{ij} + (\tfrac16
   R(\omega,e) +2\,D) \,
   B_{ij} B^{ij} \nonumber\\[.1ex]
   &\,
   - \big[\varepsilon^{ik}\,B_{ij} \,F^{+\mu\nu} \,
   R(\mathcal{V})_{\mu\nu}{}^{j}{}_{k} +\varepsilon_{ik}\,B^{ij}
   \,F^{-\mu\nu} R(\mathcal{V})_{\mu\nu j}{}^k \big] \nonumber\\[.1ex]
  &\,
  -8\, D\, \mathcal{D}^\mu A\, \mathcal{D}_\mu\bar A + \big(8\, \mathrm{i}
  R(A)_{\mu\nu} +2\, T_\mu{}^{cij}\, T_{\nu cij}\big) \mathcal{D}^\mu
  A \,\mathcal{D}^\nu\bar A  \nonumber\\[.1ex]
  &\,
  -\big[\varepsilon^{ij} \mathcal{D}^\mu T_{bc ij}\mathcal{D}_\mu
  A\,F^{+bc}+ \varepsilon_{ij} \mathcal{D}^\mu T_{bc}{}^{ij} \mathcal{D}_\mu
  \bar A\,F^{-bc}\big] \nonumber\\[.1ex]
  &\,
  -4\big[\varepsilon^{ij} T^{\mu b}{}_{ij}\,\mathcal{D}_\mu A
  \,\mathcal{D}^cF^{+}_{cb} + \varepsilon_{ij} T^{\mu
    bij}\,\mathcal{D}_\mu \bar A \,\mathcal{D}^cF^{-}_{cb}\big]
     \nonumber\\[.1ex]
    &\, + 8\, \mathcal{D}_a F^{-ab}\, \mathcal{D}^c F^+{}_{cb}  + 4\,
    F^{-ac}\, F^+{}_{bc}\, R(\omega,e)_a{}^b
     +\tfrac1{4} T_{ab}{}^{ij} \,T_{cdij} F^{-ab} F^{+cd}  \,.
\end{align}

By making various choices for the chiral multiplet~$\Phi$ that enters this formula, we can 
construct a large class of full-superspace Lagrangians. 
In our theory, we have a Weyl multiplet of weight $w=1$ and $n_{\rm v}+1$ 
vector multiplets~${\bf X}^{I}$ of weight $w=1$. Associated to each vector multiplet~${\bf X}^{I}$ is a 
reduced chiral multiplet~$\CC^{I}$. We review some relevant details in Appendix~\ref{Euclidean}.
We can build a class of Lagrangians by choosing the 
chiral multiplet~$\Phi$ above to be equal to an arbitrary holomorphic function~$f(\CC^{I})$ and 
similarly~$\bar \Phi$ to be equal to an anti-holomorphic function~$\bar g(\bar \CC^{I})$.
The weight zero conditions on~$\Phi$, $\bar \Phi$ translate to the condition that the functions~$f$, $\bar g$ are 
homogeneous functions of degree zero.
More generally, we can consider a sum of products of such functions
\be \label{genH}
\CH(\CC^{I}, \bar \CC^{I}) \= \sum_{n, \bar n} f^{(n)}(\CC^{I}) \; \bar g^{(\bar n)} (\bar \CC^{I}) \, . 
\ee

The full-superspace integral 
\be 
e^{-1} \CL \= \int d^4\theta \, d^4\bar{\theta}\; \CH(\CC^{I}, \bar \CC^{I}) 
\ee
written in components is as follows~\cite{deWit:2010za}:
\begin{align} \label{HLagcomps}
  e^{-1}\mathcal{L} =&\, \mathcal{H}_{IJ\bar K \bar L}\Big[\tfrac14
    \big( F_{ab}^-{}^I\, F^{-ab\,J}
                -\tfrac12 B_{ij}{}^I\, B^{ijJ} \big)
                \big( F_{ab}^+{}^K \, F^{+ab\,L} -\tfrac12 B^{ijK}\,
                  B_{ij}{}^L  \big)
              \nonumber\\
              & \qquad\quad +4\,\mathcal{D}_a A^I\, \mathcal{D}_b \bar A^K
                \big(\mathcal{D}^a A^J \,\mathcal{D}^b \bar A^L
                  + 2\, F^{-\,ac\,J}\,F^{+\,b}{}_c{}^L -
                  \tfrac14 \eta^{ab}\, B^J_{ij}\,B^{L\,ij}\big)
              \Big]\nonumber\\[.5ex]
   +&\,\Big\{ \mathcal{H}_{IJ\bar K}\Big[4\,\mathcal{D}_a A^I\,
     \mathcal{D}^a A^J\, \mathcal{D}^2\bar A^K
      - \big(F^{-ab\,I}\, F_{ab}^{-\,J} -\tfrac12 B^I_{ij}\, B^{Jij})
      \big( \Box_\mathrm{c} A^K + \tfrac18 F^{-\,K}_{ab}\, T^{ab ij}
          \varepsilon_{ij}\big)  \nonumber\\
  & \qquad\quad +8 \,\mathcal{D}^a A^I F^{-\,J}_{ab}
  \big( \mathcal{D}_cF^{+\,cb\,K}- \tfrac12 \mathcal{D}_c\bar A^K
              T^{ij\,cb} \varepsilon_{ij}\big) - \mathcal{D}_a
            A^I\, B^J_{ij}\,\mathcal{D}^aB^{K\,ij}\Big]
            +\mathrm{h.c.}\Big\} \+  \nonumber
\end{align}
\begin{align}
     +&\mathcal{H}_{I\bar J}\Big[ 4\big( \Box_\mathrm{c} \bar A^I + \tfrac18
         F_{ab}^{+\,I}\, T^{ab}{}_{ij} \varepsilon^{ij}\big)
     \big( \Box_\mathrm{c}  A^J + \tfrac18 F_{ab}^{-\,J}\, T^{abij}
       \varepsilon_{ij}\big) + 4\,\mathcal{D}^2 A^I \,\mathcal{D}^2
       \bar A^J \nonumber\\
       & \quad\quad +8\,\mathcal{D}_{a}F^{-\,abI\,}\,
       \mathcal{D}_cF^{+c}{}_{b}{}^J   - \mathcal{D}_a B_{ij}{}^I\,
            \mathcal{D}^a B^{ij\,J}
            +\tfrac1{4} T_{ab}{}^{ij} \,T_{cdij}
            \,F^{-ab\,I}F^{+cd\,J}
     \nonumber\\
     &\quad\quad
     +\big(\tfrac16 R(\omega,e) +2\,D\big) B_{ij}{}^I\, B^{ij\,J}   + 4\,
     F^{-ac\,I}\, F^{+}{}_{bc}{}^J \, R(\omega,e)_a{}^b  \nonumber\\
     &\quad\quad + 8\big(R^{\mu\nu}(\omega,e)-\tfrac13 g^{\mu\nu}
     R(\omega,e) +\tfrac1{4} T^\mu{}_{b}{}^{ij}\, T^{\nu b}{}_{ij}
     +\mathrm{i} R(A)^{\mu\nu} - g^{\mu\nu} D\big) \mathcal{D}_\mu A^I
     \,\mathcal{D}_\nu \bar A^J  \nonumber\\
     &\quad\quad
     - \big[\mathcal{D}_c \bar A^J \big(\mathcal{D}^c
     T_{ab}{}^{ij}\,F^{-\,I\,ab} +4
       \,T^{ij\,cb} \,\mathcal{D}^aF^{-\,I}_{ab} \big)\varepsilon_{ij}
       +[\mathrm{h.c.}; I\leftrightarrow J]  \big]\nonumber\\
     &\quad\quad -\big[\varepsilon^{ik}\, B_{ij}{}^I\, F^{+ab\,J}\,
      R(\mathcal{V})_{ab}{}^j{}_k +[\mathrm{h.c.}; I\leftrightarrow J]
      \big]  \Big] \,.
\end{align}

This can be further generalized by including the Weyl multiplet in the construction of the weight-zero 
super fields~$\Phi$, $\bar \Phi$. In this case, the corresponding function~$\CH$ is homogeneous of degree zero 
with~$\CC^{I}$ having scaling weight 1 and ${\bf W}^2$ having scaling weight 2. 
The resulting Lagrangian generalizes~\eqref{HLagcomps} with additional terms (see~Eqn.~(4.10), (4.11) in~\cite{deWit:2010za}). 
When the~${\bf W}^{2}$ multiplet is a non-zero constant, 
the additional terms drop out, and in this case the Lagrangian is proportional to~\eqref{HLagcomps}.
We shall use this fact in the next section.

\subsection{Non-renormalization of semi-classical entropy}

As reviewed in \S2, the semi-classical entropy is computed by evaluating the local effective Lagrangian of the theory
on the full-BPS solutions~\eqref{metric}, \eqref{attractor}. In addition, the first derivative of the Lagrangian 
enters the answer through the definition of the charges~\eqref{Legendre}. As we now review, all the full-superspace integrals discussed in the previous subsection, as well as their first derivatives, vanish 
when evaluated on the full-BPS configuration~\cite{deWit:2010za}.

The $AdS_2 \times S^2$ form of the metric implies
\be
R_{\mu\nu}(A) = R_{\mu\nu}{}^{ij}(\mathcal{V}) = D = R(\omega,e) = 0 \, .
\ee
The components of ${\bf W}^2$ then take the simple form (see~\eqref{Weylmult} in Appendix~\ref{Euclidean}):
\be \label{weylclass}
A|_{W^2} = (T_{ab}^{ij}\epsilon_{ij})^2 = -4w^2 \, , \qquad B_{ij}|_{W^2} = F^{-ab}|_{W^2} = C|_{W^2} = 0 \, .
\ee
In the gauge-fixed theory, when $w$ is constant, the full Weyl-squared multiplet is a constant (the lowest component is a constant, and the higher components vanish). 
It is convenient to write down the explicit values of the components of the $T$-tensor:
\be
\label{TTensor}
T^-_{ab} = \left(\begin{array}{cccc} 0 & iw & 0 & 0 \\ -iw & 0 & 0 & 0 \\ 0 & 0 & 0 & iw \\ 0 & 0 & -iw & 0 \end{array}\right) \, .
\ee

Similarly, the reduced chiral multiplet in the full-BPS configuration is also a constant. 
\be \label{constmult}
A|_{\CC^{I}} = X^I_* \, , \quad  B_{ij}|_{\CC^{I}}  = F^-_{ab}|_{\CC^{I}}  = C|_{\CC^{I}} = 0 \, . 
\ee
Now, the Lagrangian~\eqref{HLagcomps} involves only derivatives of~$A|_{\CC^{I}}$, and therefore vanishes 
on this constant solution. Similarly, as mentioned at the end of the previous subsection, the generalized Lagrangian including the contribution from the Weyl multiplet also vanishes for our solution with the Weyl and vector multiplets being constant. 

With similar arguments, the authors of~ \cite{deWit:2010za} also show that the first derivative of the Lagrangian 
with respect to all the fields vanish. From the discussion following~\eqref{Legendre}, we deduce that the charges,
and therefore the entropy, are not modified by the addition of the full-superspace integrals. 

To summarize, the full-BPS conditions imply an~$AdS_{2} \times S^{2}$ metric and constant scalar 
fields and gauge field strengths. The full-superspace integrals and their first derivatives vanish on these 
constant configurations, implying that the semi-classical black hole entropy is not modified by the inclusion
of these terms to the effective action. 
In the next section, we shall consider half-BPS solutions wherein the scalar fields are not constant and 
have a non-trivial profile in the bulk of~$AdS_{2}$.

\section{Full-superspace integrals and the quantum entropy}

Our goal is to examine the effect of the full-superspace integrals described in the previous section on the 
functional integral~\eqref{qef} for quantum black hole entropy. Using the localization technique sketched in 
\S2, we shall show now that the quantum entropy is completely insensitive to any of these full-superspace integrals.

Our method of proof is conceptually very simple. As stressed in \S2, the localizing manifold is defined using 
the off-shell supersymmetry variations and does not depend on the action. 
This means that a full-superspace integral added to the effective action can potentially affect 
the quantum entropy in \eqref{integral} in the following three ways:
\begin{enumerate}
\item It can change the value of the effective action evaluated on the localizing solutions and therefore change 
the value of $\CS_{\rm ren}$ from \eqref{Sren}.
\item It can change the measure on the localizing manifold either through the classical induced measure $[d \phi]$ or the value of the one-loop 
determinant $Z_{\rm det}$.
\item It can change the functional dependence of the electric charges $q_{I}$ on the fluctuating 
fields\footnote{The actual charges~$q_{I}$ take integer values and are fixed once and for all.}. 
(The magnetic charges $p^{I}$ are topological quantities and do not depend on the action.) 
\end{enumerate}
In~\S4.1,~\S4.2 we will discuss point 1 and we will show that all known full-superspace integrals 
that can be constructed in~$\CN=2$ supergravity at any level in the derivative expansion do not contribute 
to the renormalized action~$\CS_{\rm ren}$. Before doing so, we examine the effect on the measure, 
the one-loop determinant, and the electric charges, assuming that point~1 holds.

The classical induced measure arises from considering the localizing manifold as an embedded submanifold of the 
full field space of supergravity. It is a function of the action evaluated on the submanifold 
and of the determinant of the embedding matrix. The localizing solutions are solutions of the BPS equations 
which, in our off-shell supergravity formalism, do not change under any modification of the action. This means that 
the embedding matrix is also independent of the action.
Since, by assumption, the action evaluated on the localizing manifold does not change, the induced measure does not change\footnote{Note here that the determinant coming from the modes orthogonal to the embedding surface will
change in general, but this fact is irrelevant for our computation.} on addition of the full-superspace 
integrals. 
The one-loop determinant, by definition, is evaluated using the deformation action that is fixed once and for 
all in our first step of localization, and manifestly does not depend on the higher-derivative terms that we add to the physical action of supergravity. 

The electric charges~$q_{I}$ enter the 
functional integral in two different places, each time as a boundary term in the effective action. The first occurrence is 
the explicit coefficient of the Wilson line~\eqref{qef} which clearly does not depend on the higher-derivative action.
The other occurrence is through the boundary conditions of the gauge fields and scalar fields in the 
functional integral~\eqref{qef}. Since the boundary conditions are completely fixed by the full-BPS 
solutions~\eqref{attractor}, the charge is completely determined by the semi-classical theory, and the 
off-shell deformation inside the~$AdS_{2}$ does not affect it. We have already seen in~\S3 that the functional form of the charges 
in the semi-classical theory are not modified by the addition of full-superspace terms.

\subsection{The localizing solutions}

As described in \S2, the Weyl multiplet of the localizing solutions is fixed to its classical full-BPS value that 
was displayed explicitly in~\eqref{weylclass}. 
We now turn to the vector multiplet. For clarity, we parameterize the fluctuation away from the attractor solution by an arbitrary real 
field~$\varphi (r)$, and we shall plug back the half-BPS localizing value $\varphi (r) = \frac{C }{r}$
at the end of the computation. We have:
\be
X  = X _* + \frac{w}{4}\varphi , \quad\quad \bar{X}  = \bar{X} _* + \frac{\bar{w}}{4}\varphi  \, .
\ee
The auxiliary fields are determined by supersymmetry in terms of $\varphi$. The non-zero fields are (see Eqn.~(4.17) of~\cite{Gupta:2012cy}):
\be
Y_1^{I,1} = -Y_2^{I,2} = \frac{w\bar{w}}{8}\left((r^2-1)\partial_r\varphi  + r\varphi \right) \, . 
\ee
For $\varphi = \frac{C}{r}$, we recover our configuration~\eqref{scalars}, \eqref{auxfields} with 
$Y_1^{I,1} = -Y_2^{I,2} = \frac{w\bar{w}}{8} \frac{C^{I}}{r^{2}} \, $.

This localizing solution is extended to all the components of the reduced chiral 
multiplet~$\mathcal{C}$ following~\eqref{Cform}:
\bea
A|_{\mathcal{C} } &=& X  = X _* + \frac{w}{4} \varphi( r ) \cr
B_{ij}|_{\mathcal{C} } &=& Y_{ij}  = \varepsilon_{ik} \, \varepsilon_{jl} \, Y^{kl} \cr
F_{ab}^-|_{\mathcal{C} } &=& -\frac{\bar{w}}{16} \, T^-_{ab} \, \varphi (r) \\
C|_{\mathcal{C} } &=& -\frac{\bar{w}}{2}\mathcal{D}^2\left(\varphi ( r ) \right) + \frac{w}{64}\varphi (r )\left(T^+_{ab}\right)^2 \nn \, .
\eea
Here we made use of the fact that the superconformal d'Alembertian reduces to $\mathcal{D}^2$ for scalar fields,
and of~\eqref{attractor}. We also remind the reader that in the Euclidean continuation that we perform, the anti-chiral 
multiplet $\bar{\mathcal{C}} $ is not the complex conjugate of ${\mathcal{C}}$ (similarly, $T^-$ and $T^+$ 
are not related by complex conjugation due to our Euclidean continuation).

When $\varphi  = 0$, the half-BPS localizing configuration reduces to the full-BPS attractor solution,
and we recover the constant multiplet~\eqref{constmult}.

\subsection{Evaluation of the full-superspace Lagrangians}

As we saw in \S3, we need to build weight zero chiral multiplets to use the full-superspace formula~\eqref{KinLag}  
built out of kinetic multiplets. As a simple example, using the reduced chiral multiplet~$\CC$ associated with one vector 
multiplet~$\bf X$ and the Weyl-squared multiplet, we can build a chiral multiplet $\Phi$ of weight $w=0$ by taking the combination
\be \label{XtoPhi}
\Phi = \mathcal{C} \otimes \left({\bf W}^2\right)^{-\tfrac{1}{2}} \, .
\ee
This composite chiral superfield has the following components:
\bea \label{phimult}
A|_{\Phi } &=& \frac{1}{2iw}X _* + \frac{1}{8i}\varphi ( r ) \, ,\cr
B_{ij}|_{\Phi } &=& \frac{1}{2iw}Y_{ij}  \, , \cr
F^-_{ab}|_{\Phi } &=&  \frac{i}{32}\frac{\bar{w}}{w}T^-_{ab} \, \varphi ( r ) \, , \\
C|_{\Phi } &=& \frac{i\bar{w}}{4w}\mathcal{D}^2\left(\varphi ( r )\right) - \frac{i}{128}\left(T^+_{ab}\right)^2 \varphi ( r )\, . \nn
\eea

The kinetic Lagrangian~\eqref{KinLag} evaluated on the field configuration~\eqref{phimult} is:
\bea
\label{Lagrangianphi}
e^{-1}\mathcal{L} &=& \frac{1}{16}\mathcal{D}^2\varphi \mathcal{D}^2\varphi  + \frac{1}{8}\mathcal{D}^\mu \varphi  R(\omega,e)_\mu^{\:\:\:a} \mathcal{D}_a\varphi  + \frac{1}{16}\mathcal{D}^2\varphi \mathcal{D}^2\varphi  \cr
&&-\frac{1}{512}\varphi \mathcal{D}^2\varphi \left[\frac{w}{\bar{w}}\left(T^+_{ab}\right)^2 + \frac{\bar{w}}{w}\left(T^-_{cd}\right)^2\right] + \frac{1}{16384}\left(T^+_{ab}\right)^2\left(T^-_{cd}\right)^2\left(\varphi \right)^2 \cr
&&+\frac{w\bar{w}}{128}\partial^\mu\left[(r^2-1)\partial_r\varphi  + r\varphi \right]\partial_\mu\left[(r^2-1)\partial_r\varphi  + r\varphi \right] \\
&&+ \frac{1}{64}T^{-c}_\mu T^+_{\nu c}\mathcal{D}^\mu\varphi \mathcal{D}^\nu\varphi  -\frac{1}{64}\left[T^{+\;\mu b}T^+_{cb}\frac{w}{\bar{w}} - T^{-\;\mu b}T^-_{cb}\frac{\bar{w}}{w}\right]\mathcal{D}_\mu\varphi \mathcal{D}^c\varphi  \cr
&&-\frac{1}{128}\mathcal{D}_a\varphi  \mathcal{D}^c\varphi  T^{-ab}T^+_{cb} - \frac{1}{256}T^{-\;ac}R(\omega,e)_a^{\:\:\:b}T^+_{bc}\left(\varphi \right)^2 \cr
&&-\frac{1}{8192}\left(T^-_{ab}\right)^2\left(T^+_{cd}\right)^2\left(\varphi \right)^2 \nn \, .
\eea

The Riemann tensor of the near-horizon solution is determined completely by supersymmetry in terms of the $T^+$, $T^-$ components (see e.g.~Eqn.~(4.45) in~\cite{Mohaupt:2000mj}):
\be
R_a^{\;\;b} = \frac{1}{16}T_{ac}^-T^{+cb} \, .
\ee
Using this relation, and the explicit values of the tensor $T$ \eqref{TTensor}, the Lagrangian ~\eqref{Lagrangianphi} 
reduces to 
\be
e^{-1}\mathcal{L} = \frac{1}{8}\mathcal{D}^2\varphi \mathcal{D}^2\varphi  + \frac{1}{64}\varphi \mathcal{D}^2\varphi 
+\frac{w\bar{w}}{128}\partial^\mu\left[(r^2-1)\partial_r\varphi  + r\varphi \right]\partial_\mu\left[(r^2-1)\partial_r\varphi  + r\varphi \right] \, .
\ee
Here we have used the fact that the covariant derivative on the scalar fields reduces to the ordinary partial 
derivative. This Lagrangian can be rewritten as follows:
\be \label{Lagfin}
e^{-1}\mathcal{L} = \frac{1}{8}\mathcal{D}^2\varphi \left[r^2\mathcal{D}^2\varphi  + \frac{w\bar{w}}{8}\varphi \right] + \frac{w\bar{w}}{64}(r^2-1)\partial_r\left(r\varphi \right)\left[\mathcal{D}^2\varphi  + \frac{w\bar{w}}{32}\partial_r\left(r\varphi \right)\right] \, . 
\ee

Finally, plugging in the value $\varphi (r) = \tfrac{C^{I}}{r}$ shows that each of the two terms in the above Lagrangian vanishes, and we obtain:
\be \label{Liszero}
e^{-1}\mathcal{L} = 0 \, .
\ee

We thus have that the simplest full-superspace Lagrangian 
\be
\label{kineticintagain}
\int d^4\theta \, d^4\bar{\theta}\;\Phi \, \bar{\Phi}
\ee
for the field~$\Phi$ of~\eqref{XtoPhi} vanishes when evaluated on our localizing solutions. It is easy to check that this result also holds for a chiral field multiplied by an anti-chiral field built out of different vector multiplets:
\be
\label{kineticintIJ}
\int d^4\theta \, d^4\bar{\theta}\;\Phi^{I} \, \bar{\Phi}^{J}  \, . 
\ee
The reason is that such a Lagrangian is quadratic in the fluctuation~$\varphi$ and, when evaluated on the localizing solutions 
labelled by the real parameters~$C^{I}$, is proportional to $C^{I} C^{J}$. The $r$-dependent part
of the Lagrangian is exactly the same as in~\eqref{Lagfin} and vanishes for the same reason.

To discuss more general functions,  it is convenient to go to a gauge-fixed frame where $w$ and therefore the Weyl-squared multiplet is a constant. 
This means that the formula~\eqref{HLagcomps} for the vector multiplets that was written down for functions
of only vector multiplets can be used for functions of the vector multiplets and the Weyl-squared multiplet by 
simply replacing the weight one field~${\bf X}^{I}$ by the weight zero field~$\Phi^{I} = \mathcal{C}^{I} \otimes \left({\bf W}^2\right)^{-\tfrac{1}{2}}$. In this case, the function~$\CH$ can be an arbitrary real function~$\CH(\Phi^{I}, \bar \Phi^{I})$.
As noted below~\eqref{weylclass}, there are additional terms in the full Lagrangian, but these drop out for a constant 
Weyl multiplet, and the Lagrangian~\eqref{HLagcomps} is thus the most general Lagrangian of this type.

Our task is now clear -- we need to evaluate the Lagrangian \eqref{HLagcomps} on our 
localizing solutions~\eqref{phimult}. The Lagrangian splits into quadratic, cubic, and quartic terms 
in~$\Phi^{I}$ (and~$\bar \Phi^{I}$). The Lagrangian~\eqref{KinLag} follows 
from taking~$\CH = \Phi \, \bar \Phi$, in which case~\eqref{HLagcomps} reduces to its quadratic piece 
that vanishes on the localizing solutions as we've already seen in~\eqref{Liszero}. 
We note that the first term in the quadratic piece of~\eqref{HLagcomps}  is equal to the term $C \bar C$ in~\eqref{KinLag} by using 
the identity~\eqref{Cform}. The rest of the terms are identical.

We have already seen above that the Lagrangian~\eqref{HLagcomps} vanishes when the chiral or anti-chiral 
superfield is a constant (namely of the type~\eqref{constmult} with only the lowest component being 
non-zero and constant). 
This means that the Lagrangian evaluated on our localizing solutions is proportional to the 
fluctuations~$\varphi^{I}( r )$. Therefore, the quadratic, cubic, and quartic pieces in the Lagrangian 
are proportional to $\CH_{I\bar J} \, C^{I} \, C^{J}$, $\CH_{IJ \bar K} \, C^{I} \, C^{J} \, C^{K}$, and 
$\CH_{IJ \bar K \bar L} \, C^{I} \, C^{J} \, C^{K} \, C^{L}$ (recall that $C^{I}$ is real).
The $r$-dependent part of the Lagrangian~\eqref{HLagcomps} can therefore be extracted 
using a single superfield~$\Phi$ and its conjugate~$\bar \Phi$.

From our computation above, it is manifest that the quadratic piece vanishes on 
the full localizing solutions.  We find that the cubic part of the Lagrangian~\eqref{HLagcomps} also 
vanishes identically. The quartic term involves a subtlety 
regarding the Euclidean continuation. It contains the term $2\, F^{-\,ac\,J}\,F^{+\,b}{}_c{}^L$,
which in Minkowski signature is real since $F^-$ and $F^+$ are related by complex conjugation. 
This means that in Minkowski signature, we have
\be
2\, F^{-\,ac\,J}\,F^{+\,b}{}_c{}^L = 2\, F^{-\,ac\,J}\,{\bar F}^{-\,b}{}_c{}^L = F^{-\,ac\,J}\,{\bar F}^{-\,b}{}_c{}^L + F^{+\,ac\,J}\,{\bar F}^{+\,b}{}_c{}^L \, .
\ee
When switching to Euclidean signature, there is an ambiguity as to which formula should be continued, and we choose 
to continue the last form of the above expression. This choice guarantees that the resulting Lagrangian is 
explicitely real in Euclidean signature even though $F^-$ and $F^+$ are not related by complex conjugation
anymore. Note that this choice does not affect the continuation of the quadratic and cubic pieces of the
Lagrangian~\eqref{HLagcomps}. After performing this analytic continuation, we find that the quartic part of the Lagrangian, 
and therefore the full Lagrangian~\eqref{HLagcomps} vanishes on the localizing solutions. 
We present some details of the computation involving the cubic and quartic terms in Appendix~\ref{cubicquartic}.

\section{Discussion}

The impressive agreement between the microscopic degeneracy of states in string theory and the macroscopic quantum entropy of black holes in~$\CN=8$ string theory points to a non-renormalization theorem for the quantum entropy. From the point of view of the effective gravitational theory, it suggests that the expectation value of the Wilson line~\eqref{qef} can be computed using only a particular set of terms in the effective action. 

In the more general setting of~$\CN=2$ supergravity, we have found evidence that the Lagrangian encoded by the holomorphic prepotential function alone accounts for all the entropy, and that the presence of full-superspace integrals in the effective action do not alter this value. 

One way to prove such a non-renormalization theorem rigorously would be to analyze all the Feynman diagrams contributing to the quantum entropy in the supersymmetric~$AdS_{2}$ background that we have. 
Our approach using localization allows us to work directly with the effective action evaluated on the manifold of supersymmetric solutions. This approach naturally generalizes the method of~\cite{deWit:2010za} showing 
the non-renormalization of the semi-classical entropy to the quantum case. 

In this paper we have considered a class of full-superspace Lagrangians arising from the kinetic multiplet construction. 
To have a full proof of the non-renormalization of the quantum entropy, we should consider \emph{all} possible 
full-superspace integrals. A good way to do this may be to use manifest superspace methods and analyze all local
functionals of the superfields and their covariant derivatives. This is currently being investigated. As mentioned in \S3, there is also a more subtle point about what kind of composite superfields are allowed that one must address in order to have a complete understanding of this subject.

Another potentially interesting point is the formal parallel between our non-renormaliza{\-}tion theorem for quantum black hole entropy in~$\CN=2$ supergravity and the non-renormaliza{\-}tion theorems in~$\CN=1$ supersymmetric field 
theories~\cite{Seiberg:1994bp}. In the latter case, the principles of supersymmetry and 
holomorphy combined with the analysis of weak-coupling limits could be put together in an elegant manner
to prove such theorems. 
It would be nice if there exists a similar principle underlying the supergravity theories that we consider in this paper.

\section*{Acknowledgements}

We would like to thank Daniel Butter, Atish Dabholkar, Bernard de Wit, Stefanos Katmadas, Ivano Lodato, and Ashoke Sen for useful discussions. 
This work is supported by the ERC Advanced Grant no.~246974, {\it ``Supersymmetry: a window to non-perturbative physics''}.

\appendix

\section{Some details of the off-shell multiplets and the Euclidean continuation \label{Euclidean}}

Chiral multiplets of weight $w=1$ in superconformal gravity can be consistently reduced by imposing a constraint in superspace. In Minkowski signature, this constraint takes the form of a reality condition~\cite{deRoo:1980mm}. 
In Euclidean signature, it relates the components of a chiral multiplet to the ones of the corresponding anti-chiral multiplet:
\be
(\epsilon_{ij}\bar{D}^i \sigma_{ab} D^j)^2 \; \mathcal{C} = \mp 96 \; \square_c \bar{\mathcal{C}}\, ,
\ee
where~$\mathcal{C} = (A,\Psi_i,B_{ij},F^-_{ab},\Lambda_i,C)$ and~$\bar{\mathcal{C}} = (\bar{A},\bar{\Psi}^i,\bar{B}^{ij},F^+_{ab},\bar{\Lambda}^i,\bar{C})$ denote the chiral and anti-chiral superfields, respectively, and $\square_c$ is the superconformal d'Alembertian $D_aD^a$. We caution the reader that the bar notation above is not the usual complex conjugation but denotes the independent components of the anti-chiral multiplet, since we are in Euclidean signature. In components, this constraint reads:
\bea \label{constraintcomp}
&&B_{ij} = \pm \epsilon_{ik}\epsilon_{jl}\bar{B}^{kl}\, , \cr
&&\slashed{D} \bar{\Psi}^i = \pm \epsilon^{ij}\Lambda_j\, , \cr
&&D^b\left(F^+_{ab} \mp F^-_{ab} \pm \frac{1}{4} A T^+_{ab} - \frac{1}{4}\bar{A} T^-_{ab}\right) + \frac{3}{4}\left(\mp \bar{\chi}_i\gamma_a\Psi_j\epsilon^{ij} - \chi^i\gamma_a\bar{\Psi}^j\epsilon_{ij}\right) = 0\, ,\\
&&-2\square_c\bar{A} - \frac{1}{4}F^+_{ab}T^{+ab} - 3\bar{\chi}_i\bar{\Psi}^i \mp C = 0\, . \nonumber
\eea
where $\chi^i$ sits in the Weyl multiplet and the $T$ tensor components in Euclidean signature are given by 
\be
T^-_{ab} = \left(\begin{array}{cccc} 0 & iw & 0 & 0 \\ -iw & 0 & 0 & 0 \\ 0 & 0 & 0 & iw \\ 0 & 0 & -iw & 0 \end{array}\right), \quad T^+_{ab} = \left(\begin{array}{cccc} 0 & i\bar{w} & 0 & 0 \\ -i\bar{w} & 0 & 0 & 0 \\ 0 & 0 & 0 & -i\bar{w} \\ 0 & 0 & i\bar{w} & 0 \end{array}\right) \, .
\ee
Here we have used the following definitions for an antisymmetric tensor field $A_{\mu\nu}$ in Euclidean signature:
\be
A^{\pm}_{\mu \nu} \equiv \half \left(A_{\mu \nu} \pm \wt A_{\mu \nu} \right)\, , 
\ee
where $\wt A_{\mu\nu}$ is the Hodge dual of $A_{\mu\nu}$.

Note that the the third condition of~\eqref{constraintcomp} has the structure of a Bianchi identity (modified due to the presence of the extra fields $T$ and $\chi$ present in superconformal gravity), which means that $F_{ab}$ can be interpreted as a field strength in terms of a vector potential. When taking this vector potential to be the vector field $A_\mu^I$ sitting in the $I^{th}$ vector multiplet, this allows for an identification between the components of a chiral and a vector multiplet~\cite{deRoo:1980mm}. Defining
\be
\widehat{F}_{ab}^{\pm I} = \Big(\delta_{ab}^{\;\;\;\;cd} \pm \frac{1}{2}\epsilon_{ab}^{\;\;\;\;cd}\Big)e_c^\mu e_d^\nu\,\partial_{[\mu}A_{\nu]}^I\, , 
\ee
the identification is as follows:
\bea \label{Cform}
A|_{\mathcal{C}^I} &=& X^I \cr
\Psi_i|_{\mathcal{C}^I} &=& \Omega_i^I \cr
B_{ij}|_{\mathcal{C}^I} &=& Y_{ij}^I = \varepsilon_{ik}\varepsilon_{jl}Y^{I\:kl} \\
F_{ab}^-|_{\mathcal{C}^I} &\equiv& F_{ab}^{-I} = \widehat{F}_{ab}^{-I} + \tfrac{1}{4}\left[\bar{\psi}_\rho^{\:i}\gamma_{ab}\gamma^\rho\Omega^{I\:j} + \bar{X}^I\bar{\psi}_\rho^{\:i}\gamma^{\rho\sigma}\gamma_{ab}\psi_\sigma^{\:j} - \bar{X}^I T_{ab}^{\:\:ij}\right]\varepsilon_{ij} \cr
\Lambda_i|_{\mathcal{C}^I} &=& -\varepsilon_{ij}\slashed{D}\Omega^{I\:j} \cr
C|_{\mathcal{C}^I} &=& -2\square_c \bar{X}^I - \tfrac{1}{4}\hat{F}_{ab}^{+I} T^{+ab} - 3\bar{\chi}_i\Omega^{I\:i} \nn
\eea

In $\CN=2$, it is also possible to build another scalar chiral multiplet of weight $w=2$ by squaring the Weyl multiplet, ${\bf W}^2 = \varepsilon_{ik}\varepsilon_{jl}{\bf W}_{ab}^{\:\:ij}{\bf W}^{abkl}$. The various components are given by~\cite{Mohaupt:2000mj}:
\bea \label{Weylmult}
A|_{W^2} &=& (T_{ab}^{\:\:ij}\varepsilon_{ij})^2 \cr
\Psi_i|_{W^2} &=& 16\varepsilon_{ij}R(Q)^j_{ab}T^{-ab} \cr
B_{ij}|_{W^2} &=& -16\varepsilon_{k[i}R(\mathcal{V})^k_{\:\:j]ab}T^{-ab} - 64\varepsilon_{ik}\varepsilon_{jl}\bar{R}(Q)^k_{ab}R(Q)^{l\:ab} \cr
F^{-ab}|_{W^2} &=& -16\mathcal{R}(M)_{cd}^{\;\;\;ab}T^{-cd} - 16\varepsilon_{ij}\bar{R}(Q)^i_{cd}\gamma^{ab}R(Q)^{j\:cd} \cr
\Lambda_i|_{W^2} &=& 32\varepsilon_{ij}\gamma^{ab}R(Q)^j_{cd}\mathcal{R}(M)^{cd}_{\;\;\;ab} + 16\left(\mathcal{R}(S)_{i\:ab} + 3\gamma_{[a}D_{b]}\chi_i\right)T^{-ab} \\
&&- 64R(\mathcal{V})^{\:\:\:\:k}_{ab\:\:i}\varepsilon_{kl}R(Q)^{l\:ab} \cr
C|_{W^2} &=& 64\mathcal{R}(M)^{-cd}_{\;\;\;\;\;ab}\mathcal{R}(M)^{-ab}_{cd} + 32R(\mathcal{V})^{-ab\:k}_{\;\;\;\;\;\;\;\;\;l}\;R(\mathcal{V})^{-\:\:l}_{ab\:\:\:k} \cr
&&- 32T^{abij}D_aD^cT_{cbij} + 128\bar{\mathcal{R}}(S)^{ab}_{\:\:i}R(Q)^{\:\:i}_{ab} + 384\bar{R}(Q)^{ab\:i}\gamma_aD_b\chi_i \nn
\eea
where
\bea
R(Q)^i_{\mu\nu} &=& 2\mathcal{D}_{[\mu}\psi_{\nu]}^{\;\;i} - \gamma_{[\mu}\phi_{\nu]}^{\;\;i} - \tfrac{1}{8}T^{abij}\gamma_{ab}\gamma_{[\mu}\psi_{\nu]\;j} \cr
\mathcal{R}(M)^{\;\;\;\;ab}_{\mu\nu} &=& R(\omega,e)^{\;\;\;\;ab}_{\mu\nu} - 4f_{[\mu}^{\;\;[a}e_{\nu]}^{\;\;b]} + \tfrac{1}{2}\left(\bar{\psi}_{[\mu}^{\;\;i}\gamma^{ab}\phi_{\nu]\;i} + \textnormal{h.c.}\right) \cr
&&+ \left(\tfrac{1}{4}\bar{\psi}_{\mu}^{\;\;i}\psi_{\nu}^{\;\;j}T^{ab}_{ij} - \tfrac{3}{4}\bar{\psi}_{[\mu}^{\;\;i}\gamma_{\nu]}\gamma^{ab}\chi_i - \bar{\psi}_{[\mu}^{\;\;i}\gamma_{\nu]}R(Q)^{ab}_i + \textnormal{h.c.}\right) \\
&&+ \tfrac{1}{32}\left(T^+_{\mu\nu}T^{-ab} + T^-_{\mu\nu}T^{+ab}\right) \cr
\mathcal{R}(S)^i_{\mu\nu} &=& 2\mathcal{D}_{[\mu}\phi_{\nu]}^{\;\;i} - 2f_{[\mu}^{\;\;a}\gamma_a\psi_{\nu]}^{\;\;i} - \tfrac{3}{2}\gamma_a\psi_{[\mu}^{\;\;i}\bar{\psi}_{\nu]}^{\;\;j}\gamma^a\chi_j + \tfrac{3}{4}T_{\mu\nu}^{ij}\chi_j \nn
\eea
Note that it is also possible to build the anti-chiral multiplet $\bar{\bf W}^2$ corresponding to ${\bf W}^2$, whose lowest component is now $A|_{\bar{W}^2} = (T_{abij}\varepsilon^{ij})^2$ and a similar construction follows for the higher components.

\section{The quartic and cubic pieces of the general full-superspace Lagrangian \label{cubicquartic}}

Plugging the field values~\eqref{phimult} into the expression~\eqref{HLagcomps} leads to a differential equation on the fluctuations~$\varphi^I(r)$. The quartic piece yields: 
\bea
\frac{w^2\bar{w}^2(r^2-1)^2}{262144}\Big[& \!\!\!\!\!\!\!\!\!\!\!\!\!\!\varphi^K\left(\varphi^L + r\p_r\varphi^L\right)\Big(\varphi^I\left(\varphi^J + r\p_r\varphi^J\right) + \p_r\varphi^I\left(r\varphi^J + (r^2-1)\p_r\varphi^J\right)\Big) \\
&+\p_r\varphi^K\Big(\p_r\varphi^L\Big((r^2-1)\varphi^I \left(\varphi^J + r\p_r\varphi^J\right) + r\p_r\varphi^I\left(\varphi^J + r^2\varphi^J + r^3\p_r\varphi^J\right)\Big) \cr
&\qquad + \varphi^L\Big(r\varphi^I\left(\varphi^J + r\p_r\varphi^J\right) + \p_r\varphi^I\left((r^2+2)\varphi^J + (r^3+r)\p_r\varphi^J\right)\Big)\Big)\Big] \, , \nn
\eea
and one can check that when $\varphi^I(r) = \tfrac{C^I}{r}$, this expression reduces to 0.

For the cubic piece, we find the following expression:
\bea
-i\frac{w^2\bar{w}^2(r^2-1)}{32768} \left(\varphi^J + r\p_r\varphi^J\right) \Big[& \!\!-2\varphi^I\Big(\varphi^K + r\p_r\varphi^K\Big) + 2(2r^2-1)\p_r\varphi^K \p_r\varphi^I \cr
&+ (r^2-1) \left(-\varphi^I + r\p_r\varphi^I\right)\p_r^2\varphi^K\Big] + \rm{h.c.} \, ,
\eea
which again vanishes on our localizing configuration.

\end{document}